\RequirePackage{lineno}
\documentclass[aps,prl,twocolumn,showpacs,superscriptaddress,groupedaddress,floatfix]{revtex4}  
\usepackage{graphicx}  
\usepackage{dcolumn}   
\usepackage{bm}        
\usepackage{amssymb}   
\usepackage{epsfig}
\newcommand{\ttbar}{$t\overline{t}$}
\newcommand{\ppbar}{$p\overline{p}$}

\newcommand{\MCFM}    {{\sc{mcfm}}}
\newcommand{\PYTHIA}    {{\sc{pythia}}}
\newcommand{\ALPGEN}    {{\sc{alpgen}}}

\newcommand{\GEANT}     {{\sc{geant}}}
\newcommand{\COMPHEP}     {{\sc{comphep}}}

\newcommand{\rar}       {\rightarrow}
\newcommand{\MET}{\mbox{$\not\!\!E_T$}}

\newcommand{\invisible}[1]{}

\RequirePackage{xspace}

\newcommand{\gevcc}{\ensuremath{{\mathrm{\,Ge\kern -0.1em V\!/}c^2}}\xspace}
\newcommand{\gevc}{\ensuremath{{\mathrm{\,Ge\kern -0.1em V\!/}c}}\xspace}
\def\bc   {\begin{center}}
\def\ec   {\end{center}}

\newcommand{\explim}    {4.8}
\newcommand{\obslim}    {4.5}

\begin{document}


\hspace{5.2in} \mbox{Fermilab-Pub-10-493-E}

\title{
 Search for $\bm{WH}$ associated production in 5.3 fb$\bm{^{-1}}$ of $\bm{p\bar{p}}$~collisions at the Fermilab Tevatron\\}
\affiliation{Universidad de Buenos Aires, Buenos Aires, Argentina}
\affiliation{LAFEX, Centro Brasileiro de Pesquisas F{\'\i}sicas, Rio de Janeiro, Brazil}
\affiliation{Universidade do Estado do Rio de Janeiro, Rio de Janeiro, Brazil}
\affiliation{Universidade Federal do ABC, Santo Andr\'e, Brazil}
\affiliation{Instituto de F\'{\i}sica Te\'orica, Universidade Estadual Paulista, S\~ao Paulo, Brazil}
\affiliation{Simon Fraser University, Vancouver, British Columbia, and York University, Toronto, Ontario, Canada}
\affiliation{University of Science and Technology of China, Hefei, People's Republic of China}
\affiliation{Universidad de los Andes, Bogot\'{a}, Colombia}
\affiliation{Charles University, Faculty of Mathematics and Physics, Center for Particle Physics, Prague, Czech Republic}
\affiliation{Czech Technical University in Prague, Prague, Czech Republic}
\affiliation{Center for Particle Physics, Institute of Physics, Academy of Sciences of the Czech Republic, Prague, Czech Republic}
\affiliation{Universidad San Francisco de Quito, Quito, Ecuador}
\affiliation{LPC, Universit\'e Blaise Pascal, CNRS/IN2P3, Clermont, France}
\affiliation{LPSC, Universit\'e Joseph Fourier Grenoble 1, CNRS/IN2P3, Institut National Polytechnique de Grenoble, Grenoble, France}
\affiliation{CPPM, Aix-Marseille Universit\'e, CNRS/IN2P3, Marseille, France}
\affiliation{LAL, Universit\'e Paris-Sud, CNRS/IN2P3, Orsay, France}
\affiliation{LPNHE, Universit\'es Paris VI and VII, CNRS/IN2P3, Paris, France}
\affiliation{CEA, Irfu, SPP, Saclay, France}
\affiliation{IPHC, Universit\'e de Strasbourg, CNRS/IN2P3, Strasbourg, France}
\affiliation{IPNL, Universit\'e Lyon 1, CNRS/IN2P3, Villeurbanne, France and Universit\'e de Lyon, Lyon, France}
\affiliation{III. Physikalisches Institut A, RWTH Aachen University, Aachen, Germany}
\affiliation{Physikalisches Institut, Universit{\"a}t Freiburg, Freiburg, Germany}
\affiliation{II. Physikalisches Institut, Georg-August-Universit{\"a}t G\"ottingen, G\"ottingen, Germany}
\affiliation{Institut f{\"u}r Physik, Universit{\"a}t Mainz, Mainz, Germany}
\affiliation{Ludwig-Maximilians-Universit{\"a}t M{\"u}nchen, M{\"u}nchen, Germany}
\affiliation{Fachbereich Physik, Bergische Universit{\"a}t Wuppertal, Wuppertal, Germany}
\affiliation{Panjab University, Chandigarh, India}
\affiliation{Delhi University, Delhi, India}
\affiliation{Tata Institute of Fundamental Research, Mumbai, India}
\affiliation{University College Dublin, Dublin, Ireland}
\affiliation{Korea Detector Laboratory, Korea University, Seoul, Korea}
\affiliation{CINVESTAV, Mexico City, Mexico}
\affiliation{FOM-Institute NIKHEF and University of Amsterdam/NIKHEF, Amsterdam, The Netherlands}
\affiliation{Radboud University Nijmegen/NIKHEF, Nijmegen, The Netherlands}
\affiliation{Joint Institute for Nuclear Research, Dubna, Russia}
\affiliation{Institute for Theoretical and Experimental Physics, Moscow, Russia}
\affiliation{Moscow State University, Moscow, Russia}
\affiliation{Institute for High Energy Physics, Protvino, Russia}
\affiliation{Petersburg Nuclear Physics Institute, St. Petersburg, Russia}
\affiliation{Stockholm University, Stockholm and Uppsala University, Uppsala, Sweden }
\affiliation{Lancaster University, Lancaster LA1 4YB, United Kingdom}
\affiliation{Imperial College London, London SW7 2AZ, United Kingdom}
\affiliation{The University of Manchester, Manchester M13 9PL, United Kingdom}
\affiliation{University of Arizona, Tucson, Arizona 85721, USA}
\affiliation{University of California Riverside, Riverside, California 92521, USA}
\affiliation{Florida State University, Tallahassee, Florida 32306, USA}
\affiliation{Fermi National Accelerator Laboratory, Batavia, Illinois 60510, USA}
\affiliation{University of Illinois at Chicago, Chicago, Illinois 60607, USA}
\affiliation{Northern Illinois University, DeKalb, Illinois 60115, USA}
\affiliation{Northwestern University, Evanston, Illinois 60208, USA}
\affiliation{Indiana University, Bloomington, Indiana 47405, USA}
\affiliation{Purdue University Calumet, Hammond, Indiana 46323, USA}
\affiliation{University of Notre Dame, Notre Dame, Indiana 46556, USA}
\affiliation{Iowa State University, Ames, Iowa 50011, USA}
\affiliation{University of Kansas, Lawrence, Kansas 66045, USA}
\affiliation{Kansas State University, Manhattan, Kansas 66506, USA}
\affiliation{Louisiana Tech University, Ruston, Louisiana 71272, USA}
\affiliation{Boston University, Boston, Massachusetts 02215, USA}
\affiliation{Northeastern University, Boston, Massachusetts 02115, USA}
\affiliation{University of Michigan, Ann Arbor, Michigan 48109, USA}
\affiliation{Michigan State University, East Lansing, Michigan 48824, USA}
\affiliation{University of Mississippi, University, Mississippi 38677, USA}
\affiliation{University of Nebraska, Lincoln, Nebraska 68588, USA}
\affiliation{Rutgers University, Piscataway, New Jersey 08855, USA}
\affiliation{Princeton University, Princeton, New Jersey 08544, USA}
\affiliation{State University of New York, Buffalo, New York 14260, USA}
\affiliation{Columbia University, New York, New York 10027, USA}
\affiliation{University of Rochester, Rochester, New York 14627, USA}
\affiliation{State University of New York, Stony Brook, New York 11794, USA}
\affiliation{Brookhaven National Laboratory, Upton, New York 11973, USA}
\affiliation{Langston University, Langston, Oklahoma 73050, USA}
\affiliation{University of Oklahoma, Norman, Oklahoma 73019, USA}
\affiliation{Oklahoma State University, Stillwater, Oklahoma 74078, USA}
\affiliation{Brown University, Providence, Rhode Island 02912, USA}
\affiliation{University of Texas, Arlington, Texas 76019, USA}
\affiliation{Southern Methodist University, Dallas, Texas 75275, USA}
\affiliation{Rice University, Houston, Texas 77005, USA}
\affiliation{University of Virginia, Charlottesville, Virginia 22901, USA}
\affiliation{University of Washington, Seattle, Washington 98195, USA}
\author{V.M.~Abazov} \affiliation{Joint Institute for Nuclear Research, Dubna, Russia}
\author{B.~Abbott} \affiliation{University of Oklahoma, Norman, Oklahoma 73019, USA}
\author{B.S.~Acharya} \affiliation{Tata Institute of Fundamental Research, Mumbai, India}
\author{M.~Adams} \affiliation{University of Illinois at Chicago, Chicago, Illinois 60607, USA}
\author{T.~Adams} \affiliation{Florida State University, Tallahassee, Florida 32306, USA}
\author{G.D.~Alexeev} \affiliation{Joint Institute for Nuclear Research, Dubna, Russia}
\author{G.~Alkhazov} \affiliation{Petersburg Nuclear Physics Institute, St. Petersburg, Russia}
\author{A.~Alton$^{a}$} \affiliation{University of Michigan, Ann Arbor, Michigan 48109, USA}
\author{G.~Alverson} \affiliation{Northeastern University, Boston, Massachusetts 02115, USA}
\author{G.A.~Alves} \affiliation{LAFEX, Centro Brasileiro de Pesquisas F{\'\i}sicas, Rio de Janeiro, Brazil}
% ...
%
%
\collaboration{The D0 Collaboration\footnote{with visitors from
$^{a}$Augustana College, Sioux Falls, SD, USA,
$^{b}$The University of Liverpool, Liverpool, UK,
$^{c}$SLAC, Menlo Park, CA, USA,
$^{d}$ICREA/IFAE, Barcelona, Spain,
$^{e}$Centro de Investigacion en Computacion - IPN, Mexico City, Mexico,
$^{f}$ECFM, Universidad Autonoma de Sinaloa, Culiac\'an, Mexico,
and 
$^{g}$Universit{\"a}t Bern, Bern, Switzerland.%
}} \noaffiliation
\vskip 0.25cm
\date{December 3, 2010}

\begin{abstract}

We present a search for associated production of Higgs and $W$ bosons in $p\bar{p}$ collisions
at a center of mass energy of $\sqrt{s}=1.96$~TeV
in  5.3~fb$^{-1}$ of integrated luminosity recorded by the D0 experiment.
Multivariate analysis techniques are applied to
events containing one lepton, an imbalance in transverse energy, and one or two
$b$-tagged jets to discriminate
a potential $WH$ signal from standard model backgrounds.
We observe good agreement between data and background, and 
set an upper limit of \obslim\ (at 95\%\ confidence level and for $m_H=115$~GeV)
on the ratio of the $WH$ cross section multiplied by the branching fraction
of $H \rightarrow b \bar{b}$ to its standard model prediction.
A limit of \explim\  is expected from simulation.\\
\end{abstract}

\maketitle


The only unobserved particle of the standard model (SM) is the 
Higgs boson ($H$) which emerges from the spontaneous breaking of electroweak symmetry.
Its observation would support the hypothesis that the Higgs mechanism generates
the masses of the weak gauge bosons and accommodates
finite masses of fermions through their Yukawa couplings to the Higgs field. 
The mass of the Higgs boson ($m_H$) is not predicted by the SM, 
but the combination of direct searches at the CERN $e^+e^-$ Collider (LEP)~\cite{sm-lep}
and precision measurements of other electroweak parameters constrain $m_H$ to
$114.4<m_H<185$~GeV at the 95\% CL~\cite{elweak}.
While the region $158 < m_{H} < 175$~GeV has been excluded at the 95\%\ CL by a combination of searches at CDF and D0~\cite{ww-cdf,ww-dzero,ww-combo,tev-combo},
the remaining mass range continues to be probed at the Fermilab Tevatron Collider.
The associated production of a Higgs boson and a leptonically-decaying 
$W$ boson is among the cleanest Higgs boson search channels at the Tevatron, and 
provides the largest event yield for the decay $H \rightarrow b \bar{b}$ in the range $m_H < 135$~GeV.
Several searches for $WH$ production
at a $p\bar{p}$ center-of-mass energy of $\sqrt{s}=1.96$~TeV have been published.
Three of these~\cite{hep-ex/0410062,wh-plb,wh-prl}
use subsamples (0.17~fb$^{-1}$, 0.44~fb$^{-1}$, and 1.1 fb$^{-1}$)
of the data analyzed in this paper, while
three from the CDF collaboration are based on
cumulative samples (0.32~fb$^{-1}$,
0.95~fb$^{-1}$ and 2.7~fb$^{-1}$) 
of integrated luminosity~\cite{CDF-wh,CDF-wh-1fb,CDF-wh-2.7fb}.

We present a new search using an improved multivariate technique
based on data collected with the D0 detector, corresponding to an integrated luminosity of $5.3$~fb$^{-1}$.
The search selects events with one charged lepton ($\ell$ = electron, $e$, or muon, $\mu$), 
an imbalance in transverse energy (\MET) that arises from the unobserved
neutrino in the $W\to\ell\nu$ decay, and either two or three
jets, with one or two of these
selected as candidate $b$-quark jets ($b$-tagged).

The channels are separated into independent categories based on the 
number of $b$-tagged jets in an event (one or two). 
Single $b$-tagged events contain three important sources of backgrounds:
(i) multijet events, where a jet is misidentified as an isolated lepton, (ii)
$W$ boson production in association with $c$-quark or light-quark jets, and (iii)
$W$ boson production in association with two heavy-flavor ($b\bar{b},c\bar{c}$) jets.
In events with two $b$-tagged jets, the dominant
backgrounds are from $Wb\bar{b}$, $t\bar{t}$, and single top-quark production.


The analysis relies on  the following 
components of the D0 detector~\cite{run2det}:
(i) a central-tracking system, which consists of a silicon microstrip tracker
(SMT) and a central fiber tracker (CFT), both located within a 2~T
superconducting solenoidal magnet; 
(ii) a liquid-argon/uranium calorimeter containing electromagnetic, fine hadronic, and 
coarse hadronic layers, segmented into a central section (CC), covering 
pseudorapidity $|\eta|<1.1$ relative to the center of the detector~\cite{defs},
and two end calorimeters (EC) extending coverage to $|\eta|\approx 4.0$, all housed in 
separate cryostats~\cite{run1det}, with
scintillators between the CC and EC cryostats providing 
sampling of developing showers for $1.1<|\eta|<1.4$;
(iii) a muon system located beyond the calorimetry consisting of
layers of tracking detectors and scintillation trigger counters, 
one before and two after the 1.8~T iron toroids.  A 2006 upgrade of the D0
detector added an inner layer of silicon~\cite{layer0} to the SMT and an
improved calorimeter trigger~\cite{l1cal2b}.  The integrated luminosity is measured using 
plastic scintillator arrays located in front 
of the EC cryostats at $2.7 < |\eta| < 4.4$. 
The trigger and data acquisition systems are designed to accommodate 
high instantaneous luminosities. 

Events in the electron channel are triggered by a logical OR of several triggers 
that require an electromagnetic (EM) object or an EM object in conjunction with a jet. 
Trigger efficiencies are taken into account in the Monte Carlo (MC) simulation 
through a weighting of events based on an efficiency derived from data, and
parametrized as a function of electron $\eta$ and azimuth $\phi$, and jet transverse momentum $p_T$. 

We accept events for the muon channel from
a mixture of single high-$p_T$ muon, jet, and muon plus jet triggers, and expect 
this inclusive trigger to be fully efficient 
for our selection criteria. We verify this by comparing
events that pass a well-modeled subset of 
high-$p_T$ muon triggers to those that are selected by the inclusive set of triggers.
Good agreement is observed between data and MC for this high-$p_T$ muon subset of triggers.
Events not selected by a high-$p_T$ muon trigger tend to be selected by a jet trigger.
The efficiency of this complementary set of triggers is modeled as a function
of the scalar sum of jet $p_T$ in an event ($H_T$). This model provides a gain in efficiency 
relative to the high-$p_T$ muon triggers, and produces good agreement between data 
and MC for the combination of all triggers following its application to the simulation.


The \PYTHIA~\cite{pythia} MC generator is used to simulate
 production of dibosons with inclusive decays ($WW$, $WZ$, and $ZZ$),
$WH \rar l \nu b \bar{b}$ and $ZH \rar l l b \bar{b}$ ($l = e$, $\mu$, or $\tau$).  The contribution 
from $ZH$ events in which one lepton is not identified to the total signal yield is approximately 5\%.
Background from $W/Z$ ($V$)+jets and $t\bar{t}$ events is generated with \ALPGEN~\cite{ALPGEN}
interfaced to \PYTHIA\ for parton showering and hadronization.
The \ALPGEN\ samples are produced 
using the MLM parton-jet matching prescription~\cite{ALPGEN}.  The $V$+jets samples are divided into 
$V$+light jets and $V$+heavy-flavor jets.
The $V$+light jets samples include $Vjj$, $Vbj$, and $Vcj$ processes,
where $j$ is a light-flavor ($u,d,s$ quarks or gluons) jet, while the  $V$+heavy-flavor samples for $Vb\bar{b}$ 
and $Vc\bar{c}$ are generated separately.
Production of single top-quark events 
is generated using \COMPHEP~\cite{COMPHEP,COMPHEP2}, with \PYTHIA\ used for parton evolution and hadronization.
Simulation of both background and signal processes relies on
the CTEQ6L1~\cite{CTEQ} leading-order 
parton distribution functions for all MC events.
These events are processed through a full D0 
detector simulation based on \GEANT~\cite{GEANT}
using the same reconstruction software as used for D0 data. 
Events from randomly chosen beam crossings are overlaid on 
the simulated events to reproduce the effect of multiple $p\bar{p}$ interactions and detector noise.


The simulated background processes are normalized to their predicted SM cross sections, 
except for $W$+jets events, which are normalized to data before applying $b$-tagging,
where contamination from the $WH$ signal is expected to be negligible.  The signal cross sections and
branching fractions are calculated at next-to-next-to-leading order (NNLO) and are taken 
from Refs. \cite{signal1,signal2,signal3,signal4,signal5}, while the \ttbar\, single $t$, and diboson cross 
sections are at next-to-leading order (NLO), and taken from Ref.~\cite{ttbar_xsecs}, Ref.~\cite{stop_xsecs}, 
and the \MCFM\ program~\cite{mcfm}, respectively.  As a cross check, we compare data  with NLO
predictions for $W$+jets based on \MCFM , and find a relative data/MC normalization factor of $1.0 \pm 0.1$, 
where the normalization for data is obtained after subtracting all other expected
background processes. 
The normalizations of the $Vb\bar{b}$ and $Vc\bar{c}$ yields in MC relative
to data are consistent with the ratio of LO/NLO cross sections
predicted by \MCFM.  Therefore we apply these \MCFM\ ratios to the corresponding $W$+heavy-flavor 
and $Z$+heavy-flavor jet processes.



This analysis is based on a preselection of events with an electron of
 $p_{T}>15$~GeV, with $|\eta|<$ 1.1 or $1.5 < |\eta| < 2.5$,
or a muon of $p_{T}>15$~GeV, with $| \eta| < 1.6$. 
Preselected events are also required to have \MET$>20$~GeV, 
either two or three jets with $p_{T}>20$~GeV 
(after correction of the jet energy~\cite{jes}) and $|\eta|<2.5$,
and $H_T > 60$~GeV for 2-jet events, or $H_T > 80$~GeV for 3-jet events. 
 The \MET\ is calculated from the individual calorimeter cells in the EM
and fine hadronic layers of the calorimeter,
and is corrected for the presence of 
muons. All energy corrections to electrons and 
jets (including energy from the coarse hadronic layers associated with jets) 
are propagated into the \MET .
To suppress multijet background, events with
\mbox{$M_W^T < 40 - 0.5$ \MET}~GeV are removed, where $M_W^T$
is the transverse mass of the $W$ boson candidate. 
Events with additional charged leptons isolated from jets that pass the
flavor-dependent $p_T$ thresholds 
 $p_{T}^e>15$~GeV, $p_T^{\mu}>10$~GeV, and $p_T^{\tau}>10$ or 15~GeV, depending on $\tau$ decay channel~\cite{tauid}, 
are rejected to decrease dilepton background from $Z$ boson and $t\bar{t}$ events.
Events must have a reconstructed \ppbar\ interaction vertex (containing at least three associated tracks) that is located within $\pm40$~cm of the center of the detector in the longitudinal direction. 


Lepton candidates are identified in two steps.
In the first step, each candidate must pass 
``loose'' identification criteria.  For electrons, we require 
95\%\ of the energy in a shower to be deposited in the EM section of the calorimeter, 
isolation from other calorimeter energy deposits, spatial distributions of calorimeter
energies consistent with those expected for EM showers, and
a reconstructed track matched to the EM shower that is isolated from
other tracks.
For the ``loose'' muon, we require hits in each layer of the muon system,
scintillator hits in time with a beam crossing (to veto cosmic rays), 
a spatial match with a track in the central tracker, 
and isolation from jets within $\Delta\mathcal{R} < 0.5$ \cite{defs} to reject semileptonic decays of hadrons.
In the second step, the loose leptons are subjected to a more restrictive ``tight'' selection.
Tight electrons must satisfy more restrictive calorimeter isolation fractions and 
EM energy-fraction criteria,
and satisfy a likelihood test 
developed on $Z\to ee$ data based on eight quantities 
characterizing the EM nature of the particle interactions~\cite{ttbar-prd}. 
Tight muons must satisfy stricter isolation criteria 
on energy in the calorimeter and momenta of tracks near the
trajectory of the muon candidate. 
Inefficiencies introduced by lepton-identification and isolation criteria
are determined from $Z\to\ell\ell$ data.
The final selections rely only on events with tight leptons, with events containing only loose leptons
used to determine the multijet background.

Jets are reconstructed using a midpoint cone algorithm~\cite{blazey} with radius 0.5.
Identification requirements for jets are based on
longitudinal and transverse shower profiles, and
minimize the possibility that the jets are caused
by noise or spurious depositions of energy.
For data taken after 2006,
and in the corresponding simulation, jets must have at least two associated tracks emanating from
the reconstructed \ppbar\ interaction vertex.
Any difference in efficiency for jet identification 
between data and simulation is corrected by adjusting 
the jet energy and resolution in simulation to match those measured in data.
Comparison of \ALPGEN\ with other generators
and with data shows small discrepancies in distributions of jet pseudorapidity and
dijet angular separations~\cite{alw}.
The data are therefore used to correct the \ALPGEN\ $W$+jets and $Z$+jets MC events 
through polynomial reweighting functions parameterized by the leading and second-leading 
jet $\eta$, and $\Delta \mathcal{R}$ between the two leading jets, that bring
these distributions for the total simulated background and the 
high statistics sample of events prior to $b$-tagging
into agreement.

Instrumental background and that from semileptonic decays of hadrons, 
referred to jointly as the multijet background, are estimated from data.
The instrumental background is significant in the electron channel, where a jet with a high
EM fraction can pass electron-identification criteria, or 
a photon can be misidentified as an electron.
In the muon channel, the multijet background is less important
and arises mainly from semi-leptonic decay of heavy-flavor quarks,
where the muon passes isolation criteria.

To estimate the number of events that contain a jet that passes the
``tight'' lepton selection, we determine the probability $f_{T|L}$ 
for a ``loose'' lepton candidate, originating from a jet,
to also pass tight identification. 
This is done in events that pass preselection requirements without applying the selection
on $M_W^T$,
i.e., events that contain one loose lepton and two jets, but
 small \MET\ ($5-15$ GeV).  The total non-multijet background is
estimated from MC and subtracted from the data before estimating
the contribution from multijet events.
For electrons, $f_{T|L}$ is determined as a function of electron
$p_T$ in three regions of $|\eta|$ and four of $\Delta\phi($\MET$,e)$,
while for muons it is taken as a function of $|\eta|$ for two regions of $\Delta\phi($\MET$,\mu)$.
The efficiency for a loose lepton to pass the tight identification ($\varepsilon_{T|L}$)
is measured in $Z\to\ell\ell$ events in data, and is modeled as a function of $p_T$
for electrons and muons.
The estimation of multijet background described in Ref.~\cite{ttbar-prd}
is used to determine the multijet background directly from data, where
each event is assigned a weight that contributes to the multijet estimation
based on $f_{T|L}$ and $\varepsilon_{T|L}$ as a function of event kinematics.
Since $f_{T|L}$ depends on \MET, the scale of this estimate of the multijet background
must be adjusted when comparing to data with \MET$ > 20$ GeV.  
Before applying $b$-tagging, we perform a
fit to the $M_W^T$ distribution to set the scales for the
multijet and $W$+jets backgrounds simultaneously.



Efficient identification of $b$ jets is central to the search for $WH$ production. 
The D0 neural network (NN) $b$-tagging algorithm~\cite{NNcert}  
for identifying heavy-flavored jets is based on a combination of 
seven variables sensitive to the presence of tracks
or secondary vertices displaced significantly from the primary vertex.
All tagging efficiencies are determined separately for data and for simulated events.
We first use a low threshold on the NN output that corresponds to
 a misidentification rate of 2.7\%\ for light-flavor jets of $p_T \geq 50$~GeV
that are mistakenly tagged as heavy-flavored jets.
If two jets in an event pass this $b$-tagging requirement, the event is classified as double-$b$-tagged (DT). 
Events that are not classified as DT are considered for placement in an independent single-$b$-tag (ST)
sample, which requires exactly one jet to satisfy a more restrictive NN operating point corresponding to a 
misidentification rate of 0.9\%.
The efficiencies for identifying a jet that contains a $b$ hadron for the two NN operating 
points are $(63\pm 1)$\% and $(53\pm 1)$\%, respectively,
for a jet with a $p_T$ of 50~GeV. 
These efficiencies are determined for ``taggable''
jets, i.e., jets with at least two tracks, each with at least one hit in the SMT.
Simulated events are corrected to have the same fraction
of jets satisfying the taggability and $b$-tagging requirements as found in preselected data.

\begin{figure*}[tb]
\psfig{figure =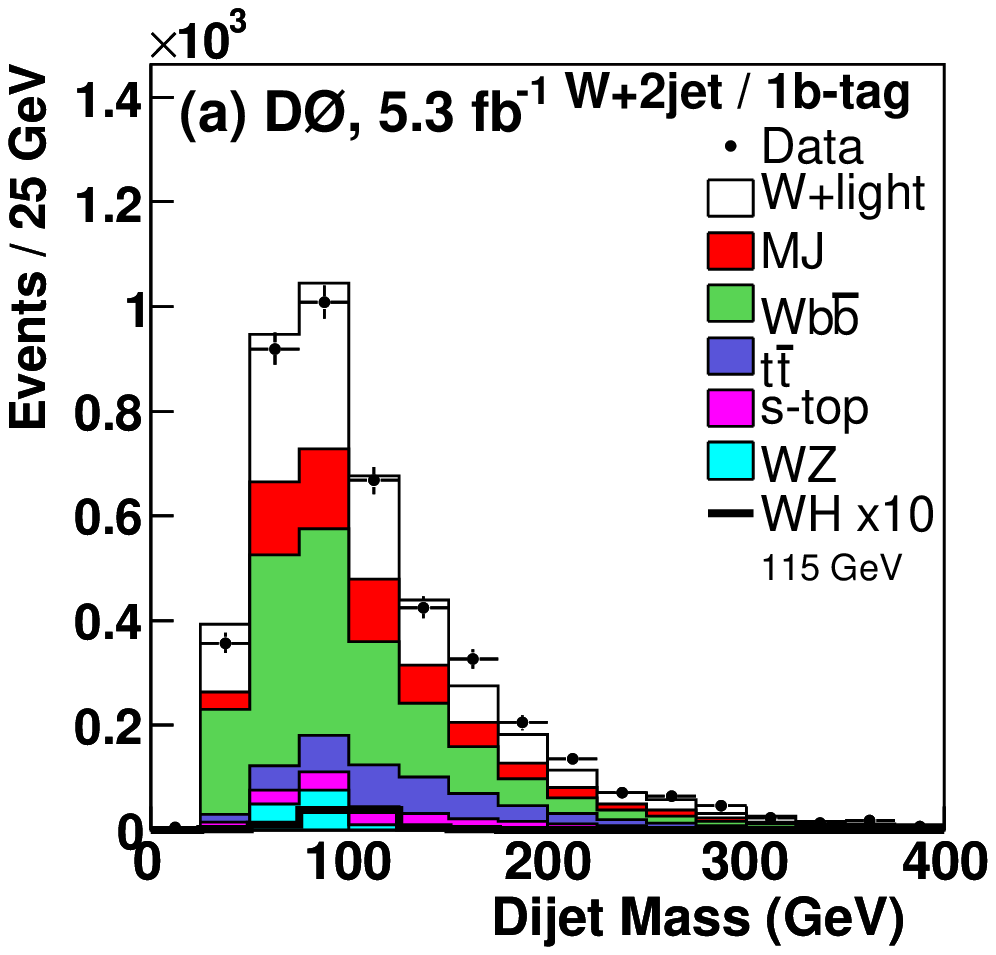,height=5.5cm}~
\psfig{figure =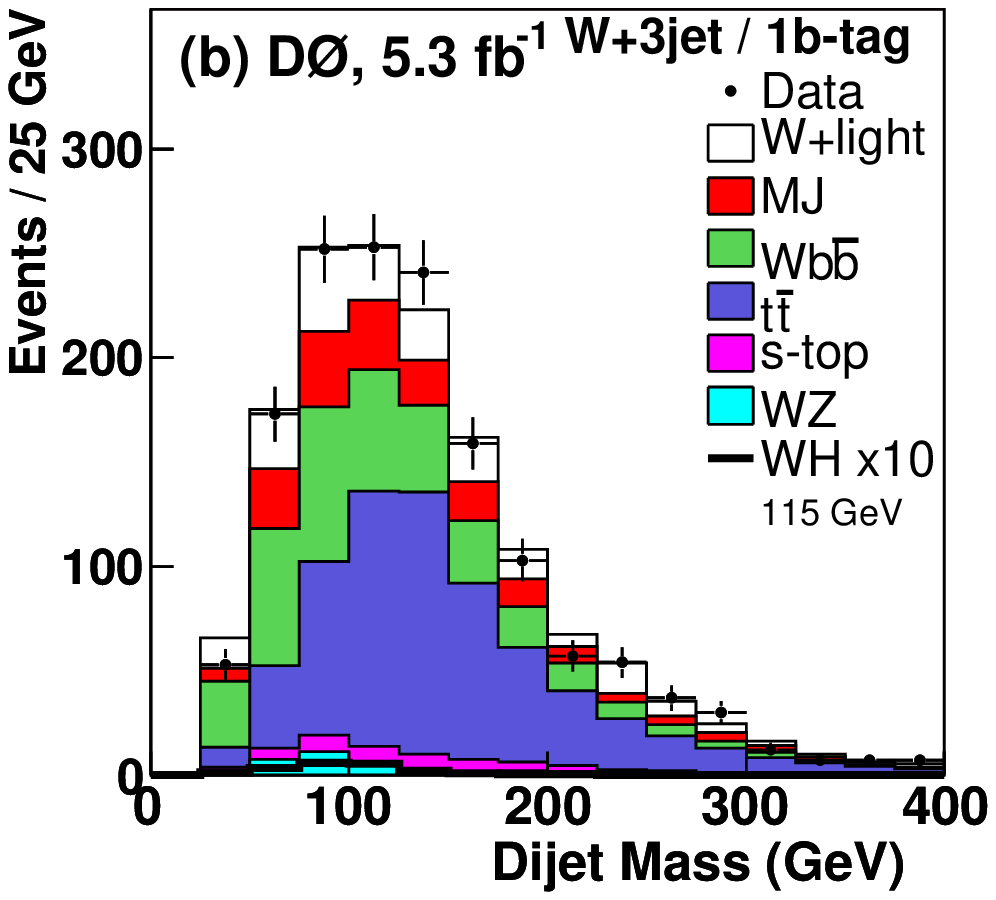,height=5.5cm}~
\psfig{figure =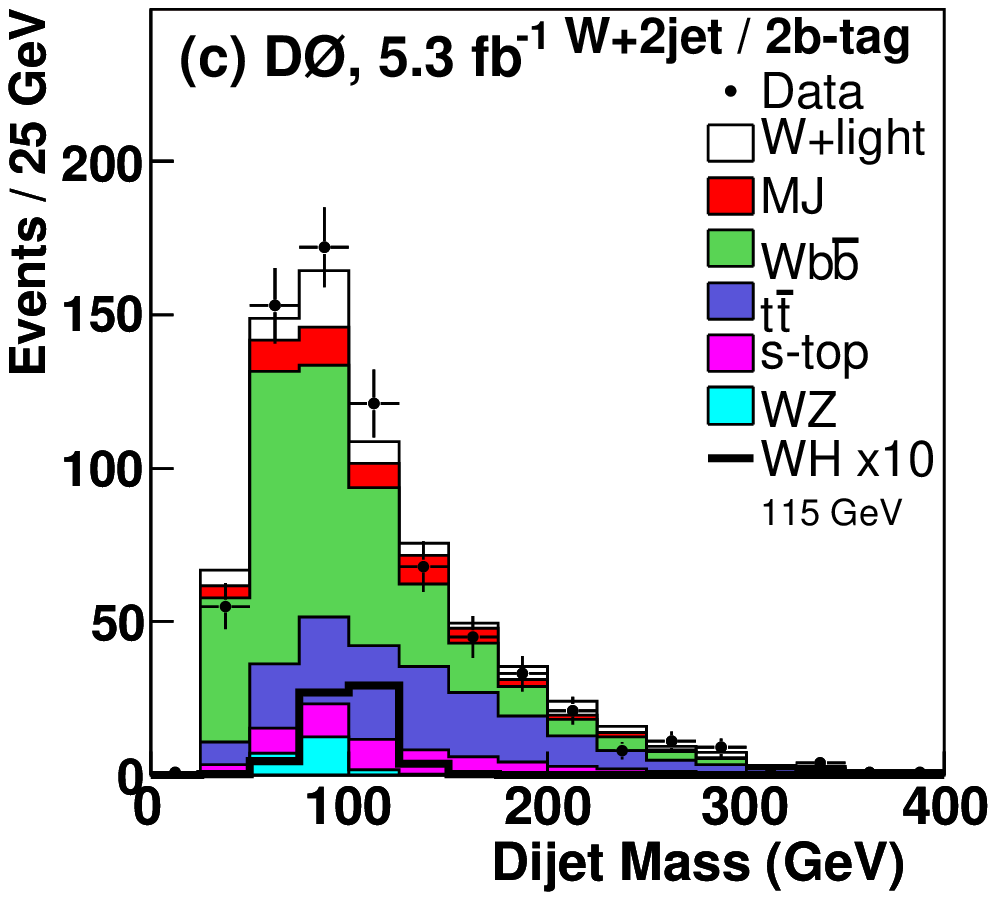,height=5.5cm}\\
\psfig{figure =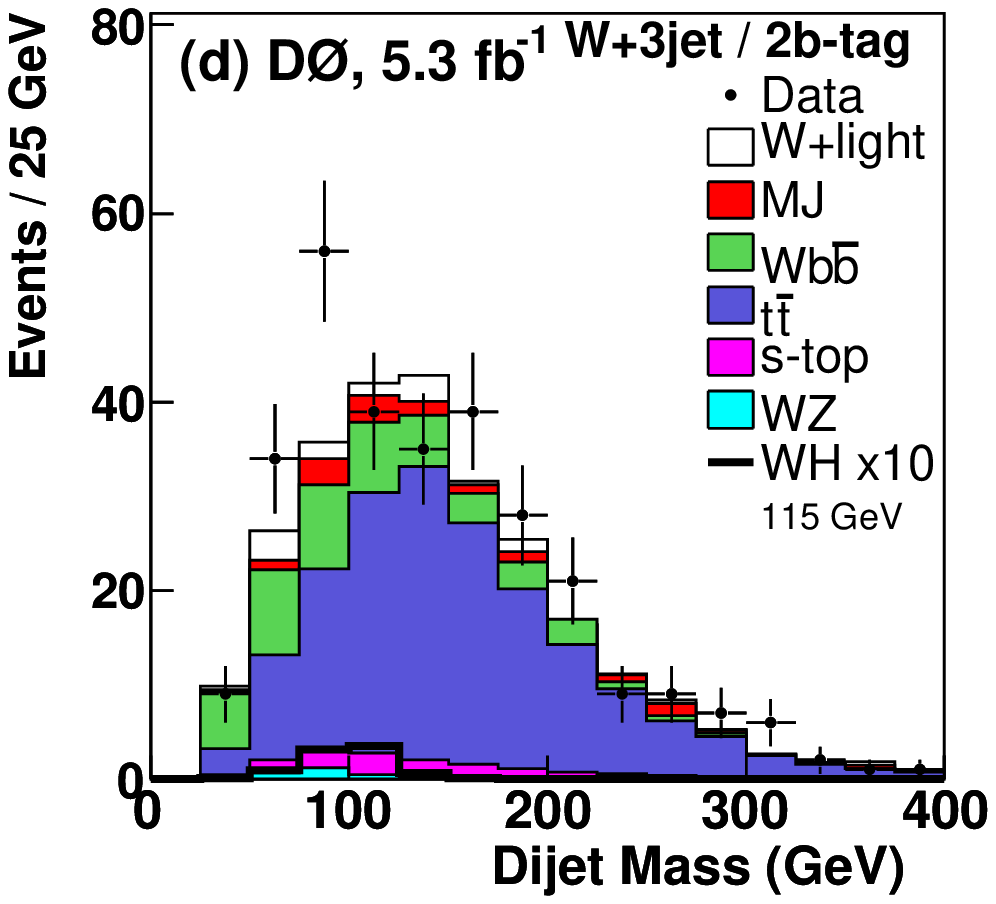,height=5.5cm}~
\psfig{figure =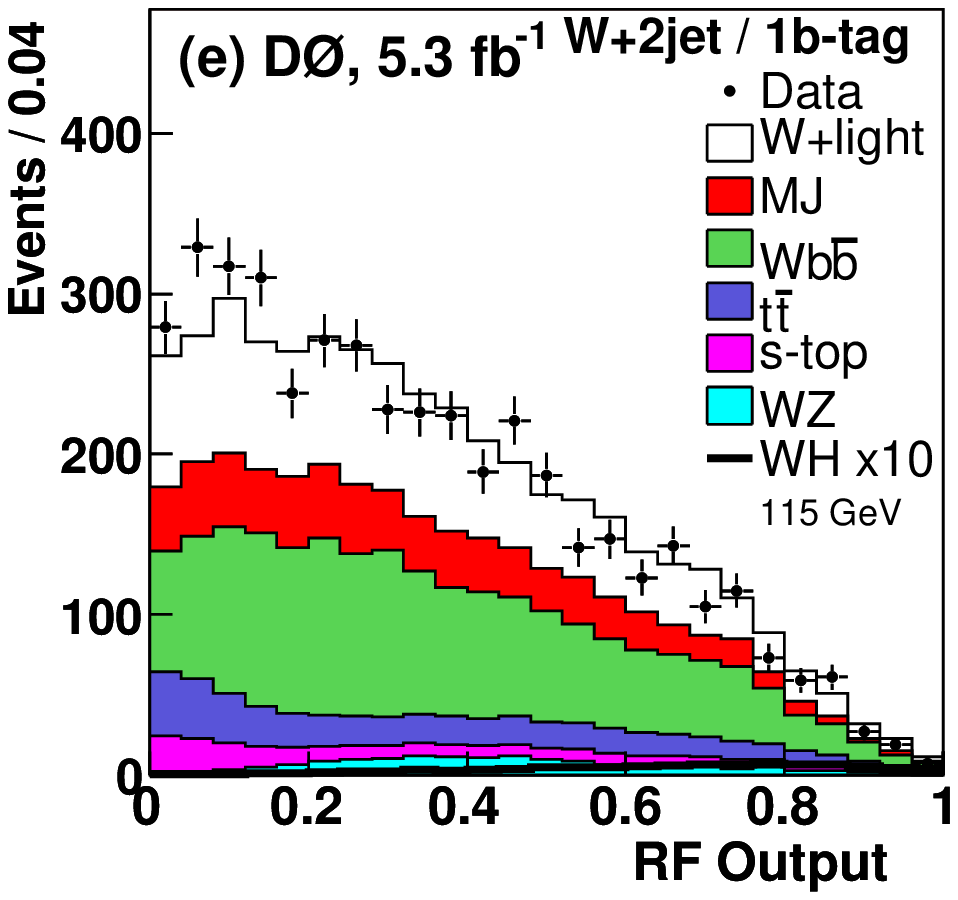,height=5.5cm}~
\psfig{figure =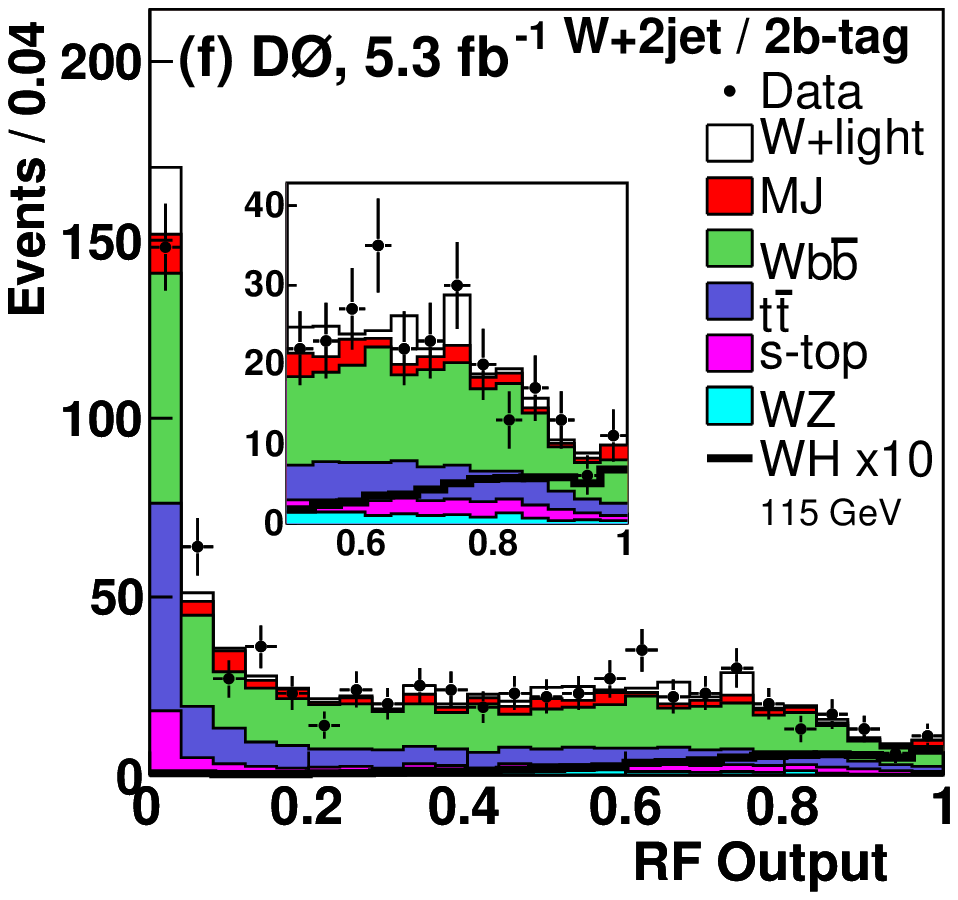,height=5.5cm}
\caption{
(Color online) Dijet mass distributions for candidate $W$-boson ST (1 $b$-tag) events with (a) 2 jets and (b) 3 jets
and for DT (2 $b$-tag) events in (c)  and (d), respectively.
The distributions in RF discriminant for 2-jet
ST and DT events, combined for lepton flavors, are shown in (e,f), respectively.
The expectation from $\sigma(p\bar{p}\to WH)\times \mathcal{B}(H\to b\bar{b})$ for $m_H = 115$ GeV is overlaid,
multiplied by a factor of 10.
}
\label{emu-two-tags}
\end{figure*}

The expected event yields following these selection criteria 
for specific backgrounds and for $m_H=115$~GeV 
are compared to the observed number of events in Table~\ref{tab:table3}.
Distributions in dijet invariant mass for the two jets of highest $p_T$,
in 2-jet and 3-jet events are shown for the
ST and DT samples in Fig.~\ref{emu-two-tags}(a--d).
The data are well-described by the sum of the simulated
SM processes and multijet background.
The contributions expected from a Higgs boson with $m_H=115$ GeV, 
multiplied by a factor of ten, are also shown for comparison.

\setlength{\unitlength}{1cm}

\begin{table}[tb]
\begin{center}
\caption{\label{tab:table3} {
Summary of event yields for the $\ell$ + $b$-tagged jets + \MET\ final state.
Event yields in data are compared with the
expected number of ST and DT events in the samples with $W$ boson candidates plus two or three jets,
comprised of contributions from simulated diboson pairs (labeled ``$WZ$'' in the table),
$W/Z$+$b\bar{b}$ or $c\bar{c}$ (``$W b\bar{b}$''), $W/Z$+light-quark jets 
(``$W+lf$''), and top-quark (``$t\bar{t}$'' and ``Single $t$'') production,
as well as data-derived multijet background (``MJ''). 
The quoted uncertainties include both
statistical and systematic contributions, including correlations between background sources and channels.
The expectation for $WH$ signal is given for $m_H=115$~GeV.
}
}
{\small
\begin{tabular}{lrclrclrclrcl}
\hline
\hline
     & \multicolumn{3}{c} { $W$+2 jet }& \multicolumn{3}{c} { $W$+2 jet }
               & \multicolumn{3}{c} { $W$+3 jet } & \multicolumn{3}{c} { $W$+3 jet }\\
     & \multicolumn{3}{c} { ST }& \multicolumn{3}{c} {  DT  }
               & \multicolumn{3}{c} { ST } & \multicolumn{3}{c} { DT }\\
\hline


$WZ$		&	153	&$\pm$&	18	&	22.5	&$\pm$&	3.3	&	33.9	&$\pm$&	4.8	&	2.6	&$\pm$&	1.1	\\
$Wb\bar{b}$	&	1601	&$\pm$&	383	&	346	&$\pm$&	93	&	358	&$\pm$&	90	&	48	&$\pm$&	13 \\
$W+lf$	        &	1290	&$\pm$&	201	&	57.5	&$\pm$&	9.2	&	210	&$\pm$&	35	&	12.1	&$\pm$&	1.8	\\
$t\bar{t}$	&	417	&$\pm$&	54	&	177	&$\pm$&	35	&	633	&$\pm$&	96	&	176	&$\pm$&	35	\\
Single $t$	&	203	&$\pm$&	33	&	58	&$\pm$&	11	&	53.6	&$\pm$&	9.1	&	13.0	&$\pm$&	2.7	\\
MJ		&	663	&$\pm$&	43	&	56.5	&$\pm$&	4.2	&	186	&$\pm$&	13	&	12.7	&$\pm$&	1.0	\\
\hline																	
All Bkg.	&	4326	&$\pm$&	501	&	718	&$\pm$&	120	&	1474	&$\pm$&	160	&	264	&$\pm$&	44	\\
$WH$		&	9.7	&$\pm$&	0.9	&	6.5	&$\pm$&	1.0	&	2.1	&$\pm$&	0.3	&	0.8	&$\pm$&	0.2	\\
Data		&	4316	&&	&		709	&&	&		1463	&&	&		301	&&		\\	
\hline    
\hline
\end{tabular}
}
\end{center}
\end{table}


We use a random forest (RF) multivariate technique~\cite{RF1,RF2} 
to separate the SM background from signal, and search for an excess, 
which is expected primarily at large values of RF discriminant.
A separate RF discriminant is used for each combination of jet 
multiplicity (two or three), lepton flavor ($e$ or $\mu$), and number of $b$-tagged jets
(one or two).  The 2-jet events are divided into data-taking periods, before and after the 2006 detector upgrade, 
for a total of twelve separately trained RFs for each chosen Higgs boson mass.  
Each RF consists of a collection of individual decision trees, with each tree considering a random subset of the 
twenty kinematic and topological input variables listed in Table \ref{tab:RFlist}.
The final RF output is the average over the individual trees.  The input variables $\sqrt{\hat{s}}$ 
and $\Delta\mathcal{R}$(dijet,$\ell+\nu$) each have two solutions arising from  the two possibilities for the 
neutrino $p_{z}$, assuming the lepton and \MET\ ($\nu$) constitute the decay products of an on-shell $W$ 
boson.   The angles $\theta^{*}$ and $\chi$ are described in  Ref.~\cite{spincorr}, and exploit 
kinematic differences arising from 
the scalar nature of the Higgs and the spins of objects in the $Wb\bar{b}$ background.  The RF outputs from 
2-jet ST and DT events are shown in Fig.~\ref{emu-two-tags}(e,f).

\begin{table}[htp]
\caption{List of RF input variables, where $j_1$ ($j_2$) refers to the jet with the highest 
(second highest) $p_T$.  \label{tab:RFlist}
}

\begin{tabular}{c|c}
\hline \hline
Variable & Definition\\
\hline
$p_{T}$($j_1$)	 	         & Leading jet $p_{T}$ \\
$p_{T}$($j_2$) 		         & Sub-leading jet $p_{T}$ \\
$E(j_{2})$  		         & Sub-leading jet energy \\
$\Delta\mathcal{R}$($j_1$,$j_2$) & $\Delta\mathcal{R}$ between jets \\
$\Delta \phi$($j_1$,$j_2$) 	 & $\Delta \phi$ between jets \\
$\Delta \phi$($j_1$, $\ell$) 	 & $\Delta \phi$ between lepton and leading jet \\
$p_{T}$(dijet system) 	 	 & $p_{T}$ of dijet system \\
$m_{jj}$ 			 & Dijet invariant mass \\
$p_{T}$($\ell$-\MET{} system) 
                                 & $p_{T}$ of $W$ candidate\\ 
\MET{} 				 & Missing transverse energy \\
aplanarity                       & See Ref.~\cite{aplan}\\
$\sqrt{\hat{s}}$ 		 & Invariant mass of the $\nu$+$\ell$+dijet system \\
\raisebox{-1.5ex}[0ex][-1.5ex]{$\Delta\mathcal{R}$(dijet,$\ell+\nu$)}
                                 & $\Delta\mathcal{R}$ between the \\
 				 &   dijet system and $\ell+\nu$ system \\
$M_W^T$ 			 & Lepton-\MET{} transverse mass  \\
\raisebox{-1.5ex}[0ex][-1.5ex]{$H_{T}$}
				 & Scalar sum of the transverse momenta \\
				 &   of all jets in the event\\
\raisebox{-1.5ex}[0ex][-1.5ex]{$H_{Z}$}
				 & Scalar sum of the longitudinal momenta \\
 				 &   of all jets in the event \\
\raisebox{-1.5ex}[0ex][-1.5ex]{$\cos \theta^*$}
				 & Cosine of angle between $W$ candidate \\
				 & and beam direction in zero-momentum frame \\
$\cos \chi$ 			 & See Ref.~\cite{ref:coschi} \\
\hline \hline
\end{tabular}
\end{table}


The dijet mass distribution is especially sensitive to $WH$ production, and was used previously to
set limits on $\sigma(p\bar{p}\to WH) \times \mathcal{B}(H \rightarrow b \bar{b})$ in Ref.~\cite{wh-plb}.
However, the gain in sensitivity using the RF output 
 as the final discriminant is about 20\%\ for a Higgs mass of 115 GeV, which, in terms of the expected limit on the $WH$ cross section, is
equivalent to a gain of about 40\%\ in integrated luminosity.


The systematic uncertainties that affect the signal and SM backgrounds  
can be categorized by the nature of their source, i.e., theoretical (e.g., uncertainty on a cross section), 
MC modeling (e.g., reweighting of \ALPGEN\ samples), or experimental (e.g., uncertainty on integrated luminosity).
Some of these uncertainties affect only the normalization of the signal or 
backgrounds, while others also affect the differential
distribution of the RF output. 

Theoretical uncertainties include uncertainties on 
the $t\bar{t}$ and single top-quark production cross sections (10\%\ and 12\%, respectively \cite{ttbar_xsecs,stop_xsecs}),
an uncertainty on the diboson production cross section (6\% \cite{mcfm}), and an uncertainty on $W$+heavy-flavor production (20\%, estimated from \MCFM).  These uncertainties
affect only the normalization of the backgrounds.  

Uncertainties from modeling that affect the distribution in the RF output include uncertainties 
on trigger efficiency as derived from data (3--5\%),
lepton identification and reconstruction efficiency (5--6\%),
reweighting of \ALPGEN\ MC samples (2\%), the MLM matching applied to $W/Z$+light-jet events ($<0.5$\%),  
and the systematic uncertainties associated with choice of renormalization and factorization 
scales in \ALPGEN\ as well as the uncertainty on the strong coupling constant (2\%).
Uncertainties on the \ALPGEN\ renormalization and factorization scales are evaluated by adjusting the nominal scale 
for each, simultaneously, by a factor of 0.5 and 2.0.  

\begin{figure}[tb]
\centerline{ \psfig{figure=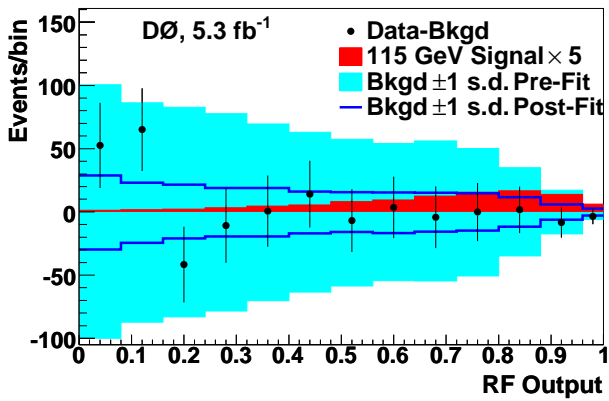,width=8.4cm} }
\caption{ \label{syst_plot}
(Color online) Distribution in the output of the RF discriminant for $m_H=115$~GeV, for the difference between data and background expectation, for all channels (both $e$ and $\mu$, ST and DT, and 2-jet and 3-jet), shown with statistical uncertainties.  
The lightly-shaded region represents the total systematic uncertainty before using constraints from data (referred to as ``Pre-Fit'' in the 
legend), while the solid lines represent the total systematic uncertainty after constraining with data (``Post-Fit'' in the legend.)  The darker shaded region represents
the SM Higgs signal expectation scaled up by a factor of 5.}
\end{figure}

Experimental uncertainties that affect only the normalization of the signal and SM backgrounds arise from the uncertainty on 
integrated luminosity (6.1\%)~\cite{luminosity_ID}.  Those that also affect the distribution in RF output include jet taggability (3\%), 
$b$-tagging efficiency (2.5--3\%\ per heavy quark-jet),
the light-quark jet misidentification rate (10\%), acceptance for jet identification (5\%); 
jet-energy calibration and resolution (varies between 15\%\ and 30\%, depending on the process 
and channel).
The background-subtracted data points for the RF discriminant for $m_H=115$~GeV,
with all channels combined, are shown with their systematic uncertainties in Fig.~\ref{syst_plot}.

\setlength{\unitlength}{1cm}

\begin{figure}[tb]
\centerline{ \psfig{figure=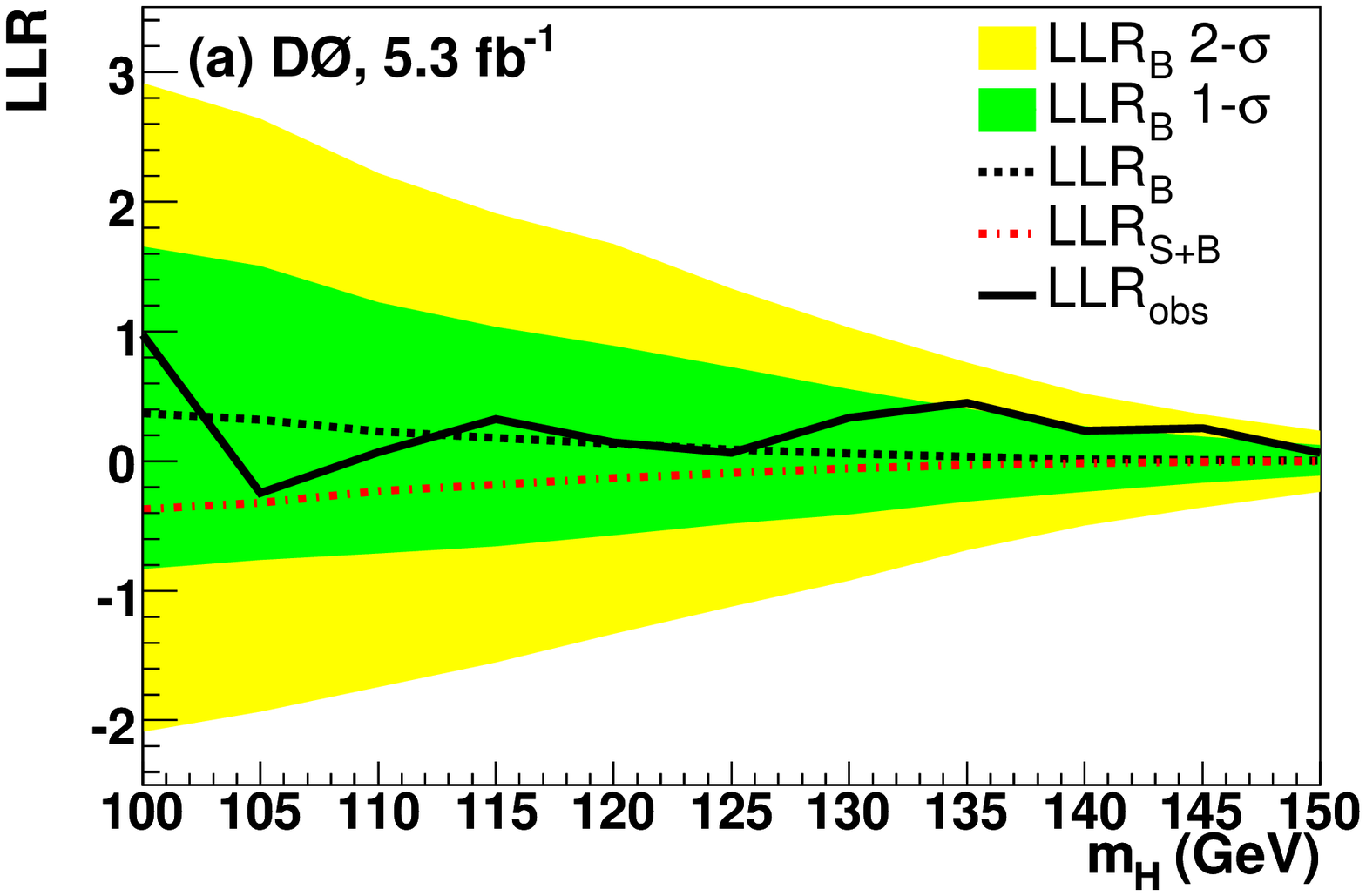,width=8.4cm} }
\centerline{ \psfig{figure=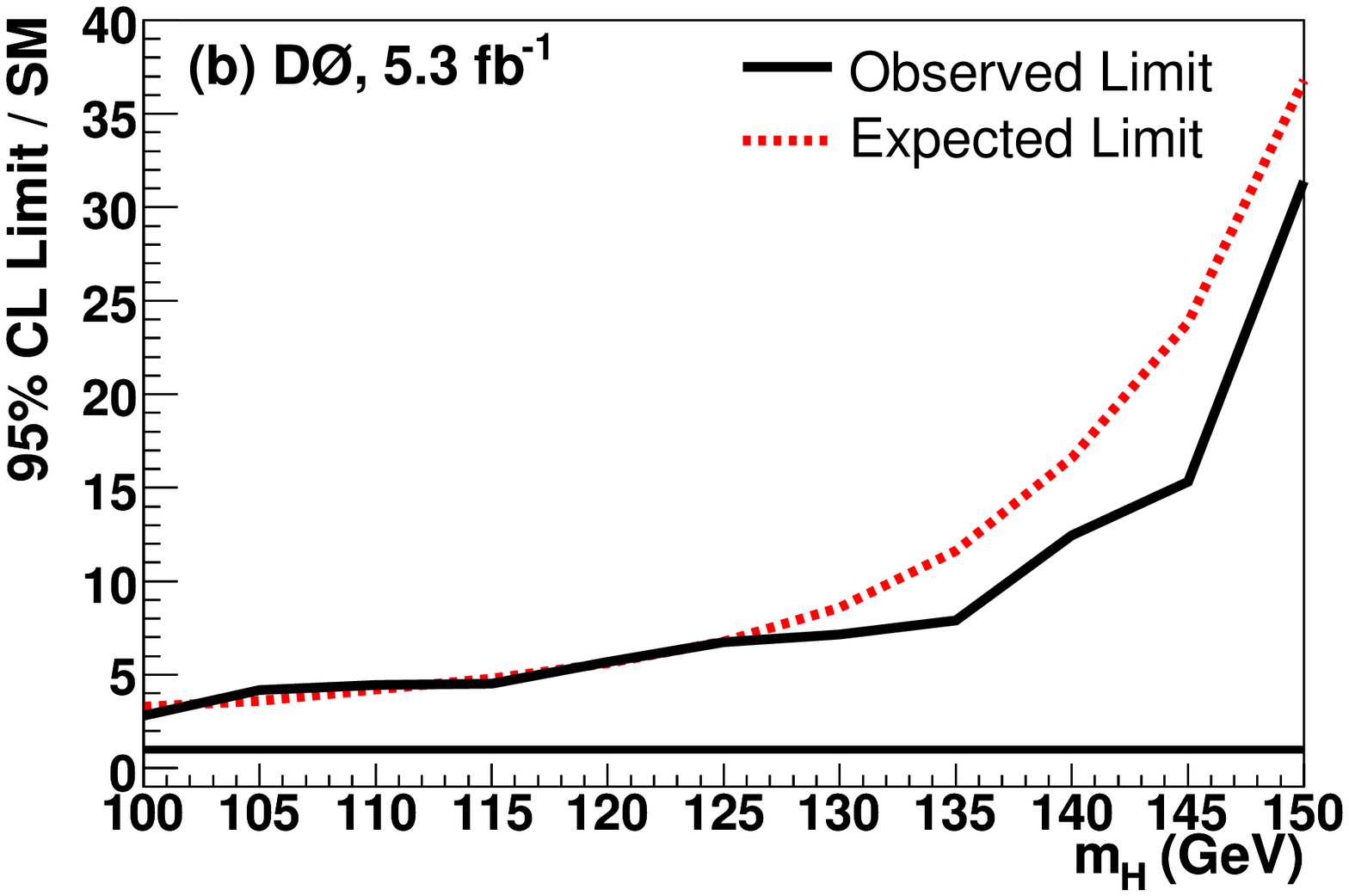,width=8.4cm} }
\centerline{
}
\caption{ \label{limits_plot} (Color online) 
(a) Log-likelihood ratios for the background-only model ($LLR_B$, with 1 and 2 standard deviation bands), 
signal$+$background model ($LLR_{S+B}$), and observation in data ($LLR_{obs}$) as a function of $m_H$.
(b) 95\% CL cross section upper limit (and corresponding expected limit from MC) on
$\sigma(p\bar{p} \rightarrow WH) \times \mathcal{B}(H \rightarrow b \bar{b})$
relative to the SM expectation, as a function of $m_H$.
}
\end{figure}

We observe no excess relative to expectation from SM background, and
we set upper limits on the production cross section $\sigma(WH)$ using
the RF outputs for the different channels.
The binning of the RF output is adjusted to assure adequate population of background events in each bin.
We calculate all limits at the 95\%\ CL using a
modified frequentist approach and a Poisson log-likelihood ratio
as test statistic~\cite{junkLim,readLim}.   The likelihood ratio is studied using pseudoexperiments based on randomly drawn 
Poisson trials of signal and background events.  
We treat systematic uncertainties 
as ``nuisance parameters'' constrained by their priors, and the best fits of these
parameters to data are determined at each value of $m_H$ by maximizing the
likelihood ratio~\cite{collie}.  Independent fits are performed to the background-only and signal-plus-background
hypotheses.  All appropriate correlations of systematic uncertainties are maintained among channels and between signal and background.
The systematic uncertainties before and after fitting are indicated in Fig.~\ref{syst_plot}. 
The log-likelihood ratios for the background-only model and the signal-plus-background model 
 as a function of  $m_H$ are shown in Fig.~\ref{limits_plot}(a).

The upper limit  on the cross section 
for $\sigma( p\bar{p} \rightarrow WH) \times \mathcal{B}(H \rightarrow b \bar{b})$
at the 95\%\ CL is a factor of \obslim\ larger than the SM expectation for $m_H=115$~GeV. 
The corresponding upper limit expected from simulation is \explim.
The analysis is repeated for ten other $m_H$ values from 100 to 150~GeV; 
the corresponding observed and expected 95\%\ CL limits relative to their SM expectations
are given in Table~\ref{limits} and in Fig.~\ref{limits_plot}(b).

\setlength{\unitlength}{1cm}

%
%
%
\begin{table*}[!tb]
\begin{center}
\caption{\label{limits}{ 
Expected and observed  95\%\ CL upper limits
on the ratio of $\sigma(p\bar{p} \rightarrow WH) \times \mathcal{B}(H \rightarrow b \bar{b})$
to its SM expectation as a function of $m_H$.}}


\begin{tabular}{p{2cm}ccccccccccc}
\hline
\hline
$m_H$ [GeV] & 100  & 105  & 110  & 115  & 120  & 125  & 130  & 135  & 140  & 145  & 150  \\
\hline
Exp. ratio  & 3.3 & 3.6 & 4.2 & 4.8 & 5.6 & 6.8 & 8.5  & 11.5  & 16.5  & 23.6  & 36.8  \\
Obs. ratio  & 2.7 & 4.0 & 4.3 & 4.5 & 5.8 & 6.6 & 7.0  &  7.6  & 12.2  & 15.0  & 30.4 \\
\hline 
\hline 
\end{tabular}
\end{center}
\end{table*}



In conclusion, $\ell+$\MET+2 or 3-jet events
have been analyzed in a search for $WH$ production in 5.3 fb$^{-1}$ of \ppbar\ collisions at the Fermilab Tevatron.
The yield of single and double $b$-tagged jets in these events is in agreement with the
expected background.  We have applied a Random Forest multivariate analysis technique to further separate signal and background.
We have set upper limits 
on $\sigma(p\bar{p} \rightarrow WH) \times \mathcal{B}(H \rightarrow b \bar{b})$
relative to their SM expectation
for Higgs masses between 100 and 150~GeV.
For $m_H=115$ GeV, the observed (expected) 95\% CL limit is a factor of \obslim\ (\explim) larger than the SM expectation.

%
We thank the staffs at Fermilab and collaborating institutions,
and acknowledge support from the
DOE and NSF (USA);
CEA and CNRS/IN2P3 (France);
FASI, Rosatom and RFBR (Russia);
CNPq, FAPERJ, FAPESP and FUNDUNESP (Brazil);
DAE and DST (India);
Colciencias (Colombia);
CONACyT (Mexico);
KRF and KOSEF (Korea);
CONICET and UBACyT (Argentina);
FOM (The Netherlands);
STFC and the Royal Society (United Kingdom);
MSMT and GACR (Czech Republic);
CRC Program and NSERC (Canada);
BMBF and DFG (Germany);
SFI (Ireland);
The Swedish Research Council (Sweden);
and
CAS and CNSF (China).
%


\begin{thebibliography}{99}

\bibitem{sm-lep} ALEPH, DELPHI, L3, and OPAL Collaborations, 
  The LEP Working Group for Higgs Boson Searches,
  Phys.\  Lett.\ B {\bf 565}, 61 (2003).

\bibitem{elweak} LEP Electroweak Working Group, \hfill\break
  {{\tt http://lepewwg.web.cern.ch/LEPEWWG/}}

\bibitem{ww-cdf} T.~Aaltonen {\em et al.} (CDF Collaborations),  Phys.\ Rev.\ Lett.\ {\bf 104}, 061803 (2010).   
\bibitem{ww-dzero} V.M.~Abazov {\em et al.} (D0 Collaboration), Phys.\ Rev.\ Lett.\ {\bf 104}, 061804 (2010).
\bibitem{ww-combo} T.~Aaltonen {\em et al.} (CDF and D0 Collaborations),  Phys.\ Rev.\ Lett.\ {\bf 104}, 061802 (2010).   
\bibitem{tev-combo}  T.~Aaltonen {\em et al.} (CDF and D0 Collaborations), FERMILAB-CONF-10-257-E (2010).

\bibitem{hep-ex/0410062} V.M.~Abazov {\em et al.} (D0 Collaboration), Phys.\ Rev.\ Lett.\ {\bf 94}, 091802 (2005).

\bibitem{wh-plb} V.M.~Abazov {\em et al.} (D0 Collaboration), Phys.\  Lett.\ B {\bf 663}, 26 (2008).

\bibitem{wh-prl} V.M.~Abazov {\em et al.} (D0 Collaboration), Phys.\ Rev.\ Lett.\ {\bf 102}, 051803 (2009).

\bibitem{CDF-wh} D.~Acosta {\em et al.} (CDF Collaboration), Phys.\ Rev.\ Lett.\ {\bf 94}, 091802 (2005).

\bibitem{CDF-wh-1fb} T.~Aaltonen {\em et al.} (CDF Collaboration), Phys.\ Rev.\ Lett.\ {\bf 100}, 041801 (2008).

\bibitem{CDF-wh-2.7fb} T.~Aaltonen {\em et al.} (CDF Collaboration), Phys.\ Rev.\ Lett.\ {\bf 103}, 101802 (2009).

\bibitem{run2det} V.M.~Abazov {\em et al.} (D0 Collaboration),  Nucl.\ Instrum.\ Methods, Phys.\ Res.\ A {\bf 565}, 463 (2006).

\bibitem{defs} Pseudorapidity $\eta = - \ln \left[ \tan\frac{\theta}{2}\right]$, 
  where $\theta$ is the polar angle as measured from the beam axis; $\phi$ is the azimuthal angle. 
  The separation between two objects in $\eta$, $\phi$ space is 
  $\Delta \mathcal{R} = \sqrt{\left(\Delta \eta\right)^2 + \left(\Delta \phi\right)^2}$.

\bibitem{run1det} S.~Abachi {\em et al.}, Nucl.\ Instrum.\ Methods Phys.\ Res.\ A { \bf 338}, 185 (1994).

\bibitem{layer0} R.~Angstadt {\em et al.},  Nucl.\ Instrum.\ Methods Phys.\ Res.\ A { \bf 622}, 298 (2010).

\bibitem{l1cal2b} M.~Abolins {\em et al.},  Nucl.\ Instrum.\ Methods Phys.\ Res.\ A { \bf 584}, 75 (2008).

\bibitem{pythia} 
         T.~Sj{\"o}strand, S.~Mrenna, and P.~Skands, J.\ High Energy Phys.\ {\bf 05}, 026 (2006), versions 6.319, 6.323 and 6.409.  Tune A was used.

\bibitem{ALPGEN} M.~Mangano {\em et al.}, J.\ High Energy Phys.\ {\bf 07}, 001 (2003), version 2.05.

\bibitem{COMPHEP} A.~Pukhov {\em et al.}, hep-ph/9908288 (1999).  

\bibitem{COMPHEP2} E.~Boos {\em et al.}, Nucl. Instrum. Meth. Phys. Res. A {\bf 534}, 250 (2004).

\bibitem{CTEQ} 
  J.~Pumplin {\em et al.}, J.\ High Energy Phys.\ {\bf 07}, 012 (2002).

\bibitem{GEANT} R.~Brun and F.~Carminati, CERN Program Library Long Writeup, Report W5013 (1993).

\bibitem{signal1} K.A.~Assamagan {\em et al.}, arXiv:hep-ph/0406152.

\bibitem{signal2} O.~Brein, A.~Djouadi, and R.~Harlander, Phys.\ Lett.\ B {\bf 579}, 149 (2004).

\bibitem{signal3} M.L.~Ciccolini, S.~Dittmaier, and M.~Kr\"amer, Phys.\ Rev.\ D {\bf 68}, 073003 (2003).

\bibitem{signal4} J.~Baglio and A.~Djouadi, J.\ High\ Energy\ Phys. {\bf 10}, 064 (2010).

\bibitem{signal5} A.~Djouadi, J.~Kalinowski and M.~Spira, Comput.\ Phys.\ Commun.\ {\bf 108}, 56 (1998).

\bibitem{ttbar_xsecs} N.~Kidonakis {\em et al.}, Phys.\ Rev.\ D {\bf 78}, 074005 (2008).

\bibitem{stop_xsecs} N.~Kidonakis, Phys.\ Rev.\ D {\bf 74}, 114012 (2006).

\bibitem{mcfm} J.M.~Campbell and R.K.~Ellis, Phys.\ Rev.\ D {\bf 60}, 113006 (1999).

\bibitem{jes} J.~Hegeman, J.\ Phys.\ Conf.\ Ser.\ {\bf 160}, 012024 (2009).

\bibitem{tauid}  V.M.~Abazov {\em et al.} (D0 Collaboration), Phys.\ Lett.\ B {\bf 670}, 292 (2009).  Only tau leptons decaying to hadrons are considered as tau candidates; those decaying to electrons or muons are included in the respective lepton contribution.


\bibitem{ttbar-prd} V.M.~Abazov {\em et al.} (D0 Collaboration), Phys.\ Rev.\ D {\bf 76}, 092007 (2007).

\bibitem{blazey}  G.~Blazey {\em et al.}, in {\sl Proceedings of the
  workshop ``QCD and Weak Boson Physics in Run II''} edited by
  U.~Baur, R.K.~Ellis, and D.~Zeppenfeld, arXiv:hep-ex/0005012, p.~47 (2000).

\bibitem{alw} J.~Alwall {\em et al.}, Eur.\ Phys.\ C {\bf 53}, 473 (2008).


\bibitem{NNcert} V.M.~Abazov {\em et al.} (D0 Collaboration), Nucl.\ Instrum.\ Methods Phys.\ Res.\ A {\bf 620}, 490 (2010).

\bibitem{RF1}  L.~Breiman, Machine Learning {\bf 45}, 5 (2001).

\bibitem{RF2}  I.~Narsky, arXiv:physics/0507143 (2005).

\bibitem{spincorr} S.~Parke and S.~Veseli, Phys.~Rev.~D {\bf 60}, 093003 (1999). 

\bibitem{aplan} Aplanarity is defined as $\frac{3}{2}\lambda_3$, where $\lambda_3$ is the smallest eigenvalue of the normalized momentum tensor \mbox{$\mathcal{M}_{ij} =\left(\sum_{o}p^{o}_{i}p^{o}_{j}\right)/\left(\sum_{o}|\vec{p}^{o}|^{2}\right)$}, where $o$ runs over the jets and charged lepton in the event, and $p^{o}_{i}$ is the $i$-th 3-momentum component of the $o$-th physics object.

\bibitem{ref:coschi} $\chi$ is the angle between the charged lepton and dijet system after boosting into the $W$ boson rest frame and then rotating the dijet system 4-vector as described in Ref.~\cite{spincorr}. 

\bibitem{luminosity_ID} T.~Andeen {\em et al}., FERMILAB-TM-2365, April 2007.

\bibitem{junkLim} T.~Junk, Nucl.\ Instrum.\ Methods Phys.\ Res.\ A {\bf 434}, 435-443 (1999).

\bibitem{readLim} A.~Read, J.\ Phys.\ G {\bf 28} 2693 (2002).

\bibitem{collie} W.~Fisher, FERMILAB-TM-2386-E (2007).

\end{thebibliography}
\end{document}